\documentclass{jpp}
\usepackage{graphicx}

\usepackage[utf8]{inputenc}
\usepackage[T1]{fontenc}
\usepackage{amsmath}
\usepackage{amsfonts}
\usepackage{float}
\usepackage{orcidlink}
\usepackage{physics,bm}
\usepackage{verbatim}

\usepackage{xcolor}
\usepackage{soul}

\usepackage{array}
\newcolumntype{C}{>{$}c<{$}} 

\newcommand{\imag}{\mathrm{i}}
\newcommand{\ca}{c_\mathrm{A}}
\newcommand{\cs}{c_\mathrm{s}}

\newcommand\scalemath[2]{\scalebox{#1}{\mbox{\ensuremath{\displaystyle #2}}}}

\allowdisplaybreaks

\shorttitle{Collisional effects on wave modes in ion-electron plasmas}
\shortauthor{De Vadder, De Jonghe \& Keppens}

\title{Collisional damping of wave modes in ion-electron plasmas}

\author{J. De Vadder\orcidlink{0009-0008-2056-9559}\aff{1}
  \corresp{\email{joeri.devadder@student.kuleuven.be}},
 J. De Jonghe \orcidlink{0000-0003-2443-3903}\aff{1,2}
  \and R. Keppens \orcidlink{0000-0003-3544-2733}\aff{1}}

\affiliation{\aff{1}Centre for mathematical Plasma Astrophysics, KU Leuven,
B-3001 Leuven, Belgium
\aff{2}School of Mathematics and Statistics, University of St Andrews, St Andrews KY16 9SS, UK}

\begin{document}

\maketitle

\begin{abstract}
To expand on recent work, we introduce collisional terms in the analysis of the warm ion-electron, two-fluid equations for a homogeneous plasma at rest. Consequently, the plasma is now described by six variables: the magnetisation, the ratio of masses over charges, the electron and ion sound speeds, the angle between the wave vector and the magnetic field, and a new parameter describing the electron-ion collision frequency. This additional parameter does not introduce new wave modes compared to the collisionless case, but does result in complex mode frequencies. Both for the backward and forward propagating modes the imaginary components are negative and thus quantify collisional damping. We provide convenient (polynomial) expressions to quantify frequencies and damping rates in all short and long wavelength limits, including the cut-off and resonance limits, whilst the one-fluid magnetohydrodynamic limit is retained with the familiar undamped slow, Alfv\'en and fast (SAF) waves. As collisions only introduce a damping, the previously introduced labelling of the wave modes S, A, F, M, O and X can be kept and assigned based on their long and short wavelength behaviour. The obtained damping at cut-off and resonance limits is parametrised with the collision frequency, and can be tailored to match known kinetic damping expressions. It is demonstrated that varying the angle can introduce crossings between the wave modes, as was already present in the ideal ion-electron case, but also a collision frequency exceeding a critical collision frequency can lead to crossings at angles where previously only avoided crossings were found.
\end{abstract}

\section{Introduction}

Two-fluid, ion-electron plasmas often appear as introductory examples in textbooks to highlight the variety of wave modes a plasma supports, such as in \citet{Thorne2017} and \citet{Gurnett2017}. The analysis of these wave modes is usually limited to propagation parallel and perpendicular to the magnetic field, which poses problems as these angles of propagation exhibit special behaviour. Properties of these wave modes have been analysed in many textbooks, e.g. \cite{Stix1992,GoedbloedKeppensPoedts2019}. They have been revisited for a variety of different conditions and angles of propagation in a number of recent works \citep{Keppens2019a,Keppens2019b,Keppens2019c,DeJonghe2020,Choi2023}. The main result from these previous works that will be generalised and re-evaluated to collisional plasma conditions is that the six ideal plasma wave modes at perpendicular and parallel orientation can exhibit crossings between the wave modes, depending on the parameter regime, whereas such crossings are strictly avoided for all oblique angles between the direction of propagation and the background magnetic field, resulting in a single frequency ordering across all wavelengths. Based on this ordering, those works emphasised a new consistent labelling scheme for all six wave mode pairs present at any angle of propagation, in ideal ion-electron plasmas at finite temperatures. Such labelling best uses the Slow-Alfv\'en-Fast (or S-A-F) labels for the lowest frequency modes, which map on the pure ideal magnetohydrodynamic (MHD) long-wavelength description, whilst the Modified Langmuir, Ordinary and eXtra-ordinary mode labels (hence M-O-X) pick up the strict ordering between generalised electrostatic and light waves at short wavelengths and high frequencies. This clear SAFMOX ordering is only broken at purely parallel or purely perpendicular orientations. At the oblique orientations, avoided crossings then signify a change in polarisation along a solution branch, as seen in e.g. \citet{Huang2019}, whereas true crossings reflect a decoupling of the two modes resulting in a stable polarisation along each solution branch.

In this paper we further explore this behaviour with the introduction of an additional parameter, the ion-electron collision frequency $\nu$. The wave modes found previously will still be present and we will therefore also use the SAFMOX labelling scheme, as presented in \citet{DeJonghe2020}, where we assign the labels based on their behaviour in the long and short wavelength limits. In all figures we will use the colour scheme green-S, red-A, blue-F, purple-M, cyan-O, black-X, as used also in our earlier studies of the collisionless variants. When branches overlap in the figures, they are drawn as dashed lines to better distinguish them.

We start from a homogeneous background with uniform magnetic field $\bm{B}$ and electrons and ions at rest. Considering the energy, momentum and continuity equations for each species $s =  i, e$ as well as the full set of Maxwell equations, we perturb with small oscillations and assume plane wave solutions $\sim \exp{\imag(\bm{k}\cdot \bm{x} -\omega t)}$. In this Fourier description, we will always adopt real-valued components of the wavevector $\bm{k}$ (such that its magnitude $k=2\pi/\lambda_w$ has a real wavelength $\lambda_w$), and in the ideal case, the frequency $\omega$ is always real-valued as well. Collisions will allow for complex-valued frequencies, which encode wave damping in their imaginary parts. Furthermore assuming charge neutrality $Z n_i = n_e$ with $Z$ the ion charge number and $n_e$, $n_i$ the electron and ion number densities respectively, a set of 8 equations in 8 variables is obtained after linearisation and some manipulation. The derivation, based on \citet{Denisse1962}, can be found in \citet{GoedbloedKeppensPoedts2019}. The result of a linearisation of these equations is also shown later in Eq. \ref{eq:dispersionmatrixfull}. The determinant of this system provides a polynomial $P(\omega, k)$ whose solutions to $P(\omega, k) = 0$ are the wave modes present in the plasma. This polynomial has successfully been used in the analysis of these wave modes in the previously mentioned works, and we will similarly start from the general system of equations. The parameters describing the plasma are the electron and ion plasma frequencies $\omega_{ps}$, cyclotron frequencies $\Omega_s$, and sound speeds $v_s$, respectively:

\begin{equation}
\begin{aligned}
    \omega_{pe} &\equiv \sqrt{\frac{e^2 n_e}{\epsilon_0 m_e}}, & \qquad \Omega_e &\equiv \frac{e B}{m_e}, & \qquad v_e &\equiv \sqrt{\frac{\gamma p_e}{n_e m_e}},\\
    \omega_{pi} &\equiv \sqrt{\frac{Z^2 e^2 n_i}{\epsilon_0 m_i}}, & \qquad \Omega_i &\equiv \frac{Z e B}{m_i}, & \qquad v_i &\equiv \sqrt{\frac{\gamma p_i}{n_i m_i}}.
\end{aligned}
\end{equation}

Here $\gamma$ denotes the ratio of specific heats, and $m_s$ and $p_s$ the mass and pressure of each species, respectively. Additionally, $e$ is the fundamental charge, $\epsilon_0$ the vacuum permittivity, and $B$ signifies the magnetic field strength $|\bm{B}|$. We furthermore introduce the plasma frequency $\omega_p \equiv \sqrt{\omega_{p e}^2 + \omega_{p i}^2}$ and combined skin depth $\delta \equiv c/\omega_p$, where $c$ is the light speed, to make all quantities dimensionless:
\begin{equation}
\begin{aligned}
    \Bar{\omega} &\equiv \omega/\omega_p, & \qquad e &\equiv \omega_{p e}/\omega_p, & \qquad E &\equiv \Omega_e/\omega_p, & \qquad v &\equiv v_e/c\\
    \Bar{k} &\equiv k \delta, & \qquad i &\equiv \omega_{p i}/\omega_p, & \qquad I &\equiv \Omega_i/\omega_p, & \qquad w &\equiv v_i/c.
\end{aligned}
\end{equation}
The remaining parameters are $\mu = Z m_e/m_i$, denoting the ratio of masses over charges, and the quantities $\lambda = \cos\theta$ and $\tau=\sin\theta$, where $\theta$ is the angle between the wave vector $\bm{k}$ and the background magnetic field. Note that
\begin{equation}
    e^2 = \frac{1}{1+\mu},\qquad i^2 = \frac{\mu}{1+\mu},\qquad I=\mu E,
\end{equation}
and the normalised sound and Alfv\'en speeds squared can be written as \citep{DeJonghe2020}
\begin{equation}
    \cs^2 = i^2v^2 + e^2w^2 = \frac{\mu v^2 + w^2}{1+\mu},\qquad \ca^2 = \frac{EI}{1+EI} = \frac{\mu E^2}{1+\mu E^2}.
\end{equation}
Finally, $\Bar{\nu} = \nu/\omega_p$ provides a measure for the electron-ion collision frequency. The collision frequency $\nu$ is related to the resistivity $\eta$ as $\eta = m_e(e^2 n_e)^{-1} \nu$. Taking for $\eta$ the Spitzer resistivity, values of $\Bar{\nu}$ are for example for typical solar coronal loop parameters of the order of $\sim 10^{-8}$. In this paper however, we will consider $\Bar{\nu}$ a free parameter and vary it across multiple orders of magnitude. This is done for two reasons: laboratory plasma conditions may indeed realise much higher collisionality, and further we will argue that a judicious choice of the collisional parameter $\Bar{\nu}$ may come from full kinetic theory, where damping rates of certain plasma wave modes are known analytically (such as is the case for collisionless Landau damping of electrostatic plasma oscillations).

Our paper is organised as follows. In Sec. \ref{sec:types_paper}, we discuss the general dispersion relation and how the electron-ion collision frequency enters. Then, we analyse various limit behaviours of the dispersion relation in Sec. \ref{sec:limits}. Lastly, we employ dispersion diagrams to illustrate the behaviour of the wave modes at parallel, perpendicular, and oblique orientations in Sec. \ref{sec:disp-diagrams} before concluding in Sec. \ref{sec:conclusion}.

\section{Dispersion relation}\label{sec:types_paper}

As noted in the introduction, the dispersion relation for a general ion-electron, two-fluid plasma with collisional effects can be obtained as the determinant of the following $8\times 8$ matrix \citep[see][]{GoedbloedKeppensPoedts2019}
\begin{align}
\setlength{\arraycolsep}{2pt}
\scalemath{0.94}{\begin{pmatrix}
&&&&&&&&\\
\omega^2 - k^2c^2 & 0 & \omega\omega_{pe} & 0 & 0 & -\omega\omega_{pi} & 0 & 0\\
&&&&&&&&\\
0 & \omega^2 - k^2c^2 & 0 & -\omega\omega_{pe} & 0 & 0 & \omega\omega_{pi} & 0\\
&&&&&&&&\\
\omega_{pe} & 0 & \omega + i\nu & \lambda\Omega_e & 0 & -i\sqrt{\mu}\nu & 0 & 0\\
&&&&&&&&\\
0 & -\omega_{pe} & \lambda\Omega_e & \omega +i\nu & \tau\Omega_e & 0 & -i\sqrt{\mu}\nu & 0\\
&&&&&&&&\\
0 & 0 & 0 & \omega \tau \Omega_e & \begin{aligned}
    &\omega^2 + i\nu\omega \\&- k^2 v_e^2 -\omega_{pe}^2
\end{aligned}& 0 & 0 & \omega_{pe}\omega_{pi} - i\sqrt{\mu}\nu\omega\\
&&&&&&&&\\
-\omega_{pi} & 0 & -i\sqrt{\mu}\nu & 0 & 0 & \omega + i\mu\nu & -\lambda\Omega_i & 0\\
&&&&&&&&\\
0 & \omega_{pi} & 0 & -i\sqrt{\mu}\nu & 0 & -\lambda\Omega_i & \omega + i\mu\nu & -\tau\Omega_i\\
&&&&&&&&\\
0 & 0 & 0 & 0 & \omega_{pe} \omega_{pi} - i\sqrt{\mu}\nu\omega & 0 & -\omega\tau\Omega_i &\begin{aligned}
    &\omega^2 + i\mu\nu\omega \\&-k^2 v_i^2 - \omega_{pi}^2
\end{aligned} \\
&&&&&&&&\\
\end{pmatrix}}
\label{eq:dispersionmatrixfull}
\end{align}
which results in a twelfth order polynomial in $\omega$ and fourth order in $k^2$ of the form
\begin{equation}
    \sum\limits_{\substack{0 \leq m \leq 6\\0 \leq n \leq 4}} \alpha_{mn}\, \bar{\omega}^{2m} \bar{k}^{2n} + \imag\bar{\nu} (1+\mu) \sum\limits_{\substack{0 \leq p \leq 5\\0 \leq q \leq 4}} \beta_{pq}\, \bar{\omega}^{2p+1} \bar{k}^{2q} = 0,
\end{equation}
where $\alpha_{mn}$ and $\beta_{pq}$ are real coefficients. Note that this implies that all even powers of $\bar{\omega}$ have a real coefficient whilst all odd powers of $\bar{\omega}$ have a purely imaginary coefficient. Just like in the ideal case, the nonzero $\alpha_{mn}$ satisfy $3 \leq m+n \leq 6$ and, similarly, we now have $3 \leq p+q \leq 5$ for the nonzero $\beta_{pq}$. This structure is highlighted in Table \ref{tab:disprel}, which shows all nonzero coefficients. Their full expressions can be found in App. \ref{app:disprel}.

\begin{table}
    \centering
    \caption{Structure of the dispersion relation. Bold coefficients are independent of $\bar{\nu}$.}
    \label{tab:disprel}
    \begin{tabular}{C|CCCCCCCCCCCCC}
        & \omega^{12} & \omega^{11} & \omega^{10} & \omega^9 & \omega^8 & \omega^7 & \omega^6 & \omega^5 & \omega^4 & \omega^3 & \omega^2 & \omega & 1 \\
        \hline
        1 & \bm{\alpha_{60}} & \bm{\beta_{50}} & \alpha_{50} & \beta_{40} & \alpha_{40} & \bm{\beta_{30}} & \bm{\alpha_{30}} &&&&&& \\
        k &&&&&&&&&&&&& \\
        k^2 &&& \bm{\alpha_{51}} & \bm{\beta_{41}} & \alpha_{41} & \beta_{31} & \alpha_{31} & \bm{\beta_{21}} & \bm{\alpha_{21}} &&&& \\
        k^3 &&&&&&&&&&&&& \\
        k^4 &&&&& \bm{\alpha_{42}} & \bm{\beta_{32}} & \alpha_{32} & \beta_{22} & \alpha_{22} & \bm{\beta_{12}} & \bm{\alpha_{12}} && \\
        k^5 &&&&&&&&&&&&& \\
        k^6 &&&&&&& \bm{\alpha_{33}} & \bm{\beta_{23}} & \alpha_{23} & \beta_{13} & \alpha_{13} & \bm{\beta_{03}} & \bm{\alpha_{03}} \\
        k^7 &&&&&&&&&&&&& \\
        k^8 &&&&&&&&& \bm{\alpha_{24}} & \bm{\beta_{14}} & \alpha_{14} & \bm{\beta_{04}} & \bm{\alpha_{04}}
    \end{tabular}
\end{table}

The presence of collisions is dictated by the collision frequency $\bar{\nu}$, which in the matrix form always appears accompanied by an imaginary factor $\imag$. In the dispersion relation itself though, both $\alpha$ and $\beta$ coefficients can feature real terms proportional to $\bar{\nu}^2$. The $\bar{\nu}$-independent coefficients have been highlighted in bold in Table \ref{tab:disprel}.

In the case of a collisionless ideal two-fluid plasma, the system can be further reduced to a $6 \times 6$ symmetric system of equations with real entries, ensuring real solutions that indicate 6 pairs of forward-backward propagating, undamped waves. In general, the governing 12th order polynomial prevents an easy factorisation, as was already noted for general ideal warm ion-electron plasmas by \citet{DeJonghe2020}. Where the resulting polynomial in the collisionless case could be written as a polynomial in $\bar{\omega}^2$ and $\bar{k}^2$, the one arising from the nonzero collision frequency is to be solved for its zeros, whilst it is written in $\bar{\omega}$ and $\bar{k}^2$. It remains possible however to obtain general information about the structure of the solutions from the polynomial.
The solutions can still be interpreted as pairs of backward and forward propagating waves, but with both solutions damped, i.e. pairs that take the form
\begin{equation}\label{eq:solutions-structure}
    \begin{cases}
        \bar{\omega}_1 = a -i b,\\
        \bar{\omega}_2 = -a -i b
    \end{cases}\qquad a,b\in\mathbb{R}^+
\end{equation}
with the symmetries $\Re(\bar{\omega}_1) = -\Re(\bar{\omega}_2)$ and $\Im(\bar{\omega}_1) = \Im(\bar{\omega}_2)$. This structure of the solutions is not unexpected compared to previous results, as the collisions only introduce a damping and with no mechanisms for growth, we should thus expect a negative imaginary part to the solutions. Furthermore the system of equations was obtained for a homogeneous medium at rest, which is not subject to effects that break the symmetry between forward and backward propagating waves. 
In the next section, we first analyse limit behaviours for the 6 wave modes, where we expect to connect with known results from corresponding ideal conditions. In order to simplify notation, we will drop the overhead bars from now on and every quantity is assumed dimensionless unless otherwise specified.

\section{Limits}\label{sec:limits}

The dispersion relation simplifies greatly when there is no external magnetic field or when one considers one of the four extremal regimes corresponding to various limits of $\omega$ and $k$. These limits are the cut-offs, resonances, local high-frequency and global low-frequency limits. Furthermore the long and short wavelength behaviour is used to assign labels to each wave mode, following the same conventions as  \citet{DeJonghe2020}.

\subsection{Unmagnetised plasma}
When the plasma is not permeated by an externally applied magnetic field, i.e. $B=0$, the parameter $E$ vanishes and consequently, so does $\lambda$. Hence, in this case the dispersion relation is only determined by $4$ parameters: $\mu$, $v$, $w$, and $\nu$. Just like in the ideal case, the dispersion relation factorises if the plasma is unmagnetised:
\begin{equation}
\begin{aligned}
    \omega^2 &\left[ \omega \left( \omega^2 - k^2 - 1 \right) + \imag\nu (1+\mu) (\omega^2 - k^2) \right]^2 \\
    &\times\left\{ \omega^4 + \imag\nu (1+\mu) \omega^3 - \omega^2 \left[ 1 + k^2 (v^2+w^2) \right] - \imag\nu (1+\mu) \omega k^2 \cs^2 + k^2 (\cs^2 + k^2 v^2 w^2) \right\} \\
    &\hspace{12cm}= 0.
\end{aligned}
\end{equation}
As expected, this equation reduces to Eq. 7 in \citet{DeJonghe2020} for an ideal plasma, i.e. $\nu = 0$. The resistive equation still features a trivial factor $\omega^2$, but the doubly degenerate factor $\omega (\omega^2-k^2-1)$ and the biquadratic factor $\omega^4 - \omega^2 \left[ 1 + k^2 (v^2+w^2) \right] + k^2 (c_\mathrm{s}^2 + k^2 v^2 w^2)$ from the ideal equation are both modified with additional terms with imaginary coefficients, proportional to $\imag\nu (1+\mu)$.

\subsection{Resonances}

The resonances are found when we look for finite asymptotic frequency $\omega$ values in the short wavelength $k \rightarrow \infty$ limit. Here we distinguish two cases, warm and cold plasmas. Indeed, our general treatment covers both these limits, as cold plasma conditions can simply set the corresponding thermal speeds $v$ and/or $w$ to zero. In the case of a warm plasma, the limit of $k \rightarrow \infty$ corresponds to considering the highest order terms in $k$, i.e. the bottom row ($\alpha_{24},\, \beta_{14},\, \alpha_{14},\, \beta_{04},\, \alpha_{04}$) in Table \ref{tab:disprel}. Dividing these terms by $v,w \neq 0$ as we are dealing with a warm plasma, the dispersion relation simplifies to the $4$th order polynomial

\begin{equation}\label{eq:limit-resonances}
    \left[ \omega^2 + \imag\nu (1+\mu) \omega - \lambda^2 \mu E^2 \right]^2 - \lambda^2 E^2 (1-\mu)^2 \omega^2 = 0.
\end{equation}

The four roots of this polynomial are then structured as in Eq. \ref{eq:solutions-structure} and correspond to the resonances found in the plasma:

\begin{equation}
    \omega (k\rightarrow\infty) = \left\{ \begin{aligned}
        &\pm \left[ \frac{1}{2} \lambda E (1-\mu) + \frac{1}{2\sqrt{2}} \sqrt{A + \sqrt{A^2+B^2}} \right] \\ &\qquad- \frac{\imag}{2} \left[ \nu (1+\mu) + \frac{1}{\sqrt{2}} \sqrt{-A + \sqrt{A^2+B^2}} \right], \\
        &\pm \left[ \frac{1}{2} \lambda E (1-\mu) - \frac{1}{2\sqrt{2}} \sqrt{A + \sqrt{A^2+B^2}} \right] \\ &\qquad- \frac{\imag}{2} \left[ \nu (1+\mu) - \frac{1}{\sqrt{2}} \sqrt{-A + \sqrt{A^2+B^2}} \right],
    \end{aligned} \right.
\end{equation}
where
\begin{equation}
    A = (\lambda^2 E^2 - \nu^2) (1+\mu)^2, \qquad B = 2\nu(1+\mu) \lambda E(1-\mu).
\end{equation}
In accordance with Eq. 13 in \citet{DeJonghe2020}, the resonances reduce to $\omega = \pm\lambda E,\, \pm \lambda\mu E$ in the collisionless case ($\nu\rightarrow 0$). We will use these expressions to predict the frequencies and their damping rates of the ion and electron cyclotron resonances found at short wavelengths, as shown further in full dispersion diagrams. 

Returning to Eq. \ref{eq:limit-resonances}, the effect of collisions becomes apparent by noting that the collision frequency always appears multiplied by the factor $(1+\mu)$. Hence, one can expect the damping of the electron and ion cyclotron resonances to depend on the ratio of masses over charges $\mu$. This is unsurprising since we expect the electrons to feel the effect of collisions with ions more than the ions due to their smaller mass. Similarly the cyclotron frequencies themselves will depend on the ratio of masses over charges as well, as could already be seen for the ideal case where the solutions differ by a factor $\mu$. As a result for a given set of parameters, we find numerically that the electron and ion resonance damping rates found in Eq. \ref{eq:limit-resonances} are approximately related by a factor $\mu$, in agreement with existing literature \citep{Braginskii1965}.

When considering the cold plasma case however, the highest order non-vanishing coefficients correspond to $k^4$ and it is from these terms that we then obtain the resonances. As there were three resonances present in the collisionless cold plasma regime \citep{Keppens2019b}, we now expect a 6th order polynomial, which turns out to be:
\begin{align}
    \nonumber& \omega^{6} + 3 \imag \nu \left(1+\mu\right) \omega^{5} - \left[1 + E^{2} \left(1 + \mu^{2}\right) + 3 \nu^{2} \left(1 + \mu\right)^{2}\right]\omega^{4}\\
    \nonumber& - \imag \nu \left(1 + \mu\right) \left[2 + \left(E^{2} + \nu^{2}\right)  \left(1 + \mu\right)^{2} \right]\omega^{3}\\
    \nonumber& +\left[\mu E^{2} (1 + \mu E^{2}) + \lambda^{2}E^{2}\left(1-\mu+\mu^2\right) + \nu^{2}\left(1 + \mu\right)^2 (1 + 2\mu E^2) \right]\omega^2\\
    \nonumber& +\imag \nu \left(1 + \mu\right) \mu E^{2} \left(1 + \lambda^2 + \mu E^{2}\right)\omega - \lambda^2 \mu^2 E^{4} = 0 .\\
\end{align}
This reduces to Eq. 10 in \citet{Keppens2019b} for $\nu\rightarrow 0$, as expected. One could use this polynomial expression valid for cold ion-electron plasmas to quantify numerically how the three resonances behave at varying angles (i.e. $\lambda$), for different magnetisations $E$, or study the special case of a cold pair plasma where $\mu=1$.

\subsection{Cut-offs}

The cut-offs are found where the frequency values attain finite limit values in the long wavelength limit $k \rightarrow 0$. This corresponds to only considering the coefficients independent of $k$ in the general dispersion relation, i.e. the top row ($\alpha_{60},\, \beta_{50},\, \alpha_{50},\, \beta_{40},\, \alpha_{40},\, \beta_{30},\, \alpha_{30}$) in Table \ref{tab:disprel}, and leads to the following polynomial describing the cut-offs:

\begin{align}\label{eq:cut-offs}
    \omega^6 \left[ \omega^2 + \imag\nu (1+\mu) \omega - 1 \right] \left[ \left( \omega^2 + \imag\nu (1+\mu)\omega - \mu E^2 -1 \right)^2 - (1-\mu)^2 E^2 \omega^2 \right] = 0.
\end{align}

Again only the odd powers of $\omega$ have imaginary coefficients that scale with $\nu$ so that in the limit $\nu \rightarrow 0$ the polynomial reduces to the one considered in the collisionless case \citet{Keppens2019b}. Where in the purely ideal case a total of three solutions in $\omega^2$ were expected, we now find $6$ solutions in $\omega$, which appear in three pairs of the form given by Eq.~\ref{eq:solutions-structure}, showing that all three cut-off frequency limits are changed by a (collisional) damping mechanism.

Denoting
\begin{equation}
    C = 4 + (E^2-\nu^2) (1+\mu)^2, \qquad D = 2\nu (1+\mu) E(1-\mu),
\end{equation}
the solutions of Eq. \ref{eq:cut-offs} can be written as
\begin{equation}
    \omega (k\rightarrow 0) = \left\{ \begin{aligned}
        &\pm \frac{1}{2} \sqrt{4 - \nu^2 (1+\mu)^2} - \frac{1}{2} \imag\nu (1+\mu), \\
        \omega_\mathrm{u} \equiv &\pm \left[ \frac{1}{2} E (1-\mu) + \frac{1}{2\sqrt{2}} \sqrt{C + \sqrt{C^2+D^2}} \right] \\ &\qquad- \frac{\imag}{2} \left[ \nu (1+\mu) + \frac{1}{\sqrt{2}} \sqrt{-C + \sqrt{C^2+D^2}} \right], \\
        \omega_\mathrm{l} \equiv &\pm \left[ \frac{1}{2} E (1-\mu) - \frac{1}{2\sqrt{2}} \sqrt{C + \sqrt{C^2+D^2}} \right] \\ &\qquad- \frac{\imag}{2} \left[ \nu (1+\mu) - \frac{1}{\sqrt{2}} \sqrt{-C + \sqrt{C^2+D^2}} \right].
    \end{aligned} \right.
\end{equation}
Note that these expressions are consistent with the collisionless limit $\nu\rightarrow 0$ found in Eq. 7 of \citet{Keppens2019b}. In this limit, the first expression reduces to $\pm 1$, whereas the second and third solution reduce to the square roots of the expression for the upper and lower cut-off frequencies $\omega_\mathrm{u}^2$ and $\omega_\mathrm{l}^2$ there, respectively, as already suggested by the notation.

These limits are influenced by only two parameters (apart from $\nu$), namely the magnetisation $E$ and the charge to mass ratio $\mu$, and do not depend on the propagation angle of the wave. We will use the complex zeros of Eq. \ref{eq:cut-offs} to check cut-off limits attained at long wavelengths in the dispersion diagrams that follow. Note that similarly with the collisionless case the upper limit $\omega_\mathrm{u}$ always corresponds to the X mode, whereas the M and O mode are defined by their behaviour in the local high-frequency limit. As a result, assigning a mode label to each cut-off frequency depends on whether the M and O modes cross or not. This depends on the parameter regime \citep{DeJonghe2020}.

\subsection{Local, high-frequency limit}

Similar to the collisionless case, we consider both $\omega, k \rightarrow \infty$ whilst keeping $\omega/k$ finite. In Table \ref{tab:disprel}, this means that the terms on the bottom diagonal ($\alpha_{60},\, \alpha_{51},\, \alpha_{42},\, \alpha_{33},\, \alpha_{24}$) become dominant. However, since these terms are independent of $\nu$, keeping only these terms in the leading order approximation, as was done for the ideal case by \citet{DeJonghe2020}, discards all the information about the wave damping. Therefore, including the lower $\beta$-diagonal ($\beta_{50}$, $\beta_{41}$, $\beta_{32}$, $\beta_{23}$, $\beta_{14}$) as well provides a first order correction to the undamped limit solution. This results in a dispersion relation of terms of combined order $12$ and $11$ in $\omega$ and $k$. Dividing by $k^{12}$ and adopting the notation $y = \omega/k$ gives, in factorised form,

\begin{equation}\label{eq:local-high-freq-col}
    y^3 (y^2-1)^2 \left\{ y (y^2-v^2) (y^2-w^2) + \frac{\imag\nu (1+\mu)}{k} \left[ 2(y^2-v^2) (y^2-w^2) + y^2 (y^2-\cs^2) \right] \right\} = 0
\end{equation}
As anticipated, the electromagnetic X and O modes $\omega^2/k^2 = 1$ travel at light speed and are undamped, whereas the last factor, describing the electron and ion sound wave limits, i.e. $\omega^2/k^2 = v^2$ and $\omega^2/k^2 = w^2$ respectively in the ideal case, is modified by a collisional term, indicative of damping. Further note that the equation does not feature $E$ and the waves are thus unaffected by the external magnetic field in this limit. Next, observe that the equation simplifies even more if one considers cold ions ($w=0$) or electrons ($v=0$).

Finally, note that the factor describing the sound waves is actually fifth order in $y$, i.e. a pair of forward-backward travelling sound waves, one for each species, and one extra mode. This remaining solution is an evanescent wave with $\Re(\omega) = 0$ and $\Im(\omega) < 0$. Since this is likely an artefact of the imperfect approximation and the mode does not propagate, it will also not be illustrated in the following plots.

Again setting $\nu = 0$, the polynomial simplifies greatly and reduces to

\begin{equation}
    y^4 (y^2-1)^2 (y^2-v^2) (y^2-w^2) = 0,
\end{equation}
which is in line with the collisionless limit of Eq. 14 in \citet{DeJonghe2020}.
This local, high-frequency limit is relevant for magneto-ionic theory, where they typically use the (cold) Appleton-Hartree equation, as well as for its warm extension \citep{DeJonghe2021a}, or to quantify the effect of Faraday rotation on electromagnetic waves \citep{Keppens2019b, DeJonghe2021a}.

\subsection{Global, low-frequency limit}

The global, low-frequency limit found by taking $\omega, k \rightarrow 0$ whilst keeping $\omega/k$ finite, corresponding to the upper diagonal ($\alpha_{30},\, \alpha_{21},\, \alpha_{12},\, \alpha_{03}$) in Table \ref{tab:disprel}, remains unaltered with the introduction of an electron-ion collisional term since ideal MHD can be directly obtained from the linearised two-fluid, ion-electron model, where this collisional effect vanishes due to conservation of momentum. As a result the exact solution obtained in \citet{DeJonghe2020} can be quoted directly (albeit with the substitution $I = \mu E$):
\begin{equation}
    \cfrac{\omega^2}{k^2} \rightarrow
    \begin{cases}
        \lambda^2 \ca^2,\\
        \cfrac{1}{2(1+ \mu E^2)}\big\{\mu E^2 + c_s^2 + \lambda^2 \mu E^2 c_s^2 \pm \big[\lambda^4 \mu^2 E^4 c_s^4\\
        \qquad\qquad\quad + 2 \lambda^2 \mu E^2 c_s^2 (c_s^2- \mu E^2-2) + (\mu E^2 + c_s^2)^2 \big]^{1/2} \big\}.
    \end{cases}
\end{equation}

As noted there, this is actually the relativistically correct expression for the familiar Alfv\'en, slow and fast magnetoacoustic MHD wave branches, which are undamped in a homogeneous plasma as studied here. This limit behaviour is one of the main motivations for introducing the SAFMOX labelling scheme. The various limits known from MHD theory (to cold conditions where $c_s=0$, or to the Newtonian regime) were also mentioned in our study of warm pair plasmas (where $E=I$ and $\mu=1$) in \cite{Keppens2019c}. Note that the ideal MHD wave limits show clear anisotropic behaviour as usually visualised in Friedrichs diagrams, and bring out that the S and A branches (slow and Alfv\'en) always vanish for perpendicular propagation ($\lambda=0$).

\section{Dispersion diagrams}\label{sec:disp-diagrams}

The dispersion diagrams can now be obtained by solving for the complex eigenfrequencies $\omega$ that appear as zeros for the determinant of Eq. \ref{eq:dispersionmatrixfull} for a given set of the parameters $E, v, w$ and $\mu$ whilst varying $k$. Each such diagram is furthermore defined by the angle parameter $\lambda$ (and hence $\tau$) and collision frequency $\nu$. As the typical parameter values are small and appear in powers when calculating the coefficients, the computation of the solutions is performed using arbitrary precision arithmetic. We illustrate the wave modes in two cases, comparing with previous works, the cold pair plasma case discussed in \cite{Keppens2019a} and warm ion-electron plasmas covered in \cite{DeJonghe2020}, starting with the cold pair plasma case.

In the collisionless $\nu=0$ cases, the six branches exhibited different behaviour for parallel, perpendicular and oblique orientations, with wave mode crossings present in the parallel and perpendicular cases and avoided crossings for all oblique angles. The crossings at purely parallel and purely perpendicular orientations are well-known, as it is in these special cases that the polynomial factorises easily, and this has been used throughout the literature to discuss those cases as representative. However, the SAFMOX labelling intends to acknowledge the strict frequency ordering at all oblique orientations instead. The same three orientations will again be investigated for a nonzero collision frequency, paying attention to crossings of previously avoided crossings and vice versa. Since the solutions exhibit the symmetries expressed in Eq. \ref{eq:solutions-structure}, it suffices to focus on only one of the solutions in each pair of wave modes. From now on, we work with the forward travelling modes ($\mathrm{Re}(\omega) > 0$). In each figure, the upper graph illustrates the real parts (i.e. the temporal frequency) of the solutions and the lower graph the imaginary parts (i.e. the damping rate). Furthermore, dots shown at the end of all curves are used to illustrate the predicted limits obtained in the previous section. 

Cold pair plasmas are obtained from Eq. \ref{eq:dispersionmatrixfull} by taking $\mu = 1$ and $v$, $w = 0$. This greatly simplifies the system and reduces the number of wave modes present in the plasma, as indeed there is no more Slow MHD wave at all, so only the AFMOX branches survive. For warm ion-electron plasmas no further simplifications are made and the full system of equations is used, and all SAFMOX branches exist. The only difference between the warm ion-electron plasmas discussed in \citet{DeJonghe2020} and those analysed here is the presence of a nonzero collision frequency $\nu$.

\subsection{Parallel Propagation}

\begin{figure}
    \centering
    \includegraphics[width=1.1\textwidth]{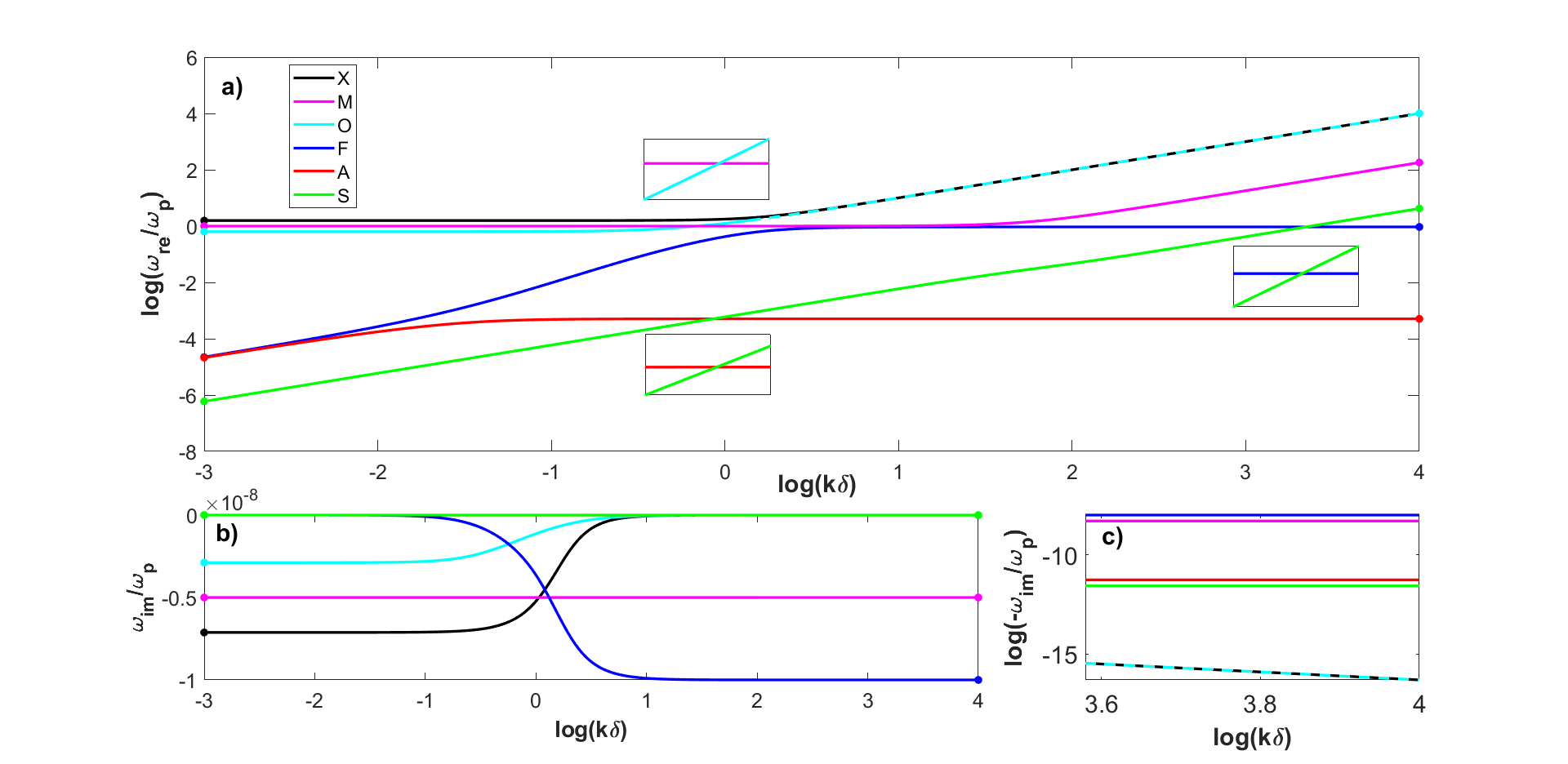}
    \caption{Dispersion diagram for parallel propagation for typical coronal loop parameters, taking $\mu \approx 1/1836$, $E = 0.935$, $v = 0.018$, $w = 0.0004$ and $\nu = 10^{-8}$. a) Wave frequencies, with insets showing the crossing behaviour. b) Damping rates. c) Logarithmic scale of damping rates in small wavelength regime.}
    
    \label{fig:coronal-n=1e-8-0g}
\end{figure}

For a value of $\lambda = 1$ corresponding to parallel propagation, Fig. \ref{fig:coronal-n=1e-8-0g} illustrates the typical behaviour of the six wave modes. This figure is for a warm solar coronal loop plasma, with dimensionless collision frequency $\nu=10^{-8}$ (all parameters used are in the caption). The wave modes behave similarly with the collisionless case: the behaviour of the real frequency variation of the wave modes is not altered nor do previously present crossings become avoided crossings when varying values of $\nu$. For this set of parameters and purely parallel propagation, Fig.~\ref{fig:coronal-n=1e-8-0g} shows that three crossings are present, the OM, AS and SF crossings. The effect of the finite collisional damping is shown clearly for all branches in the bottom panel of Fig.~\ref{fig:coronal-n=1e-8-0g} (note the scale: all dampings have order $10^{-8}$ in correspondence with the value for $\nu$), where it is seen that the three cut-off limits show clear damping in agreement with our prediction from Eq.~\ref{eq:cut-offs}, whilst the F and M branches also have finite damping. 

\subsection{Perpendicular Propagation}

\begin{figure}
    \centering
    \includegraphics[width=1.1\textwidth]{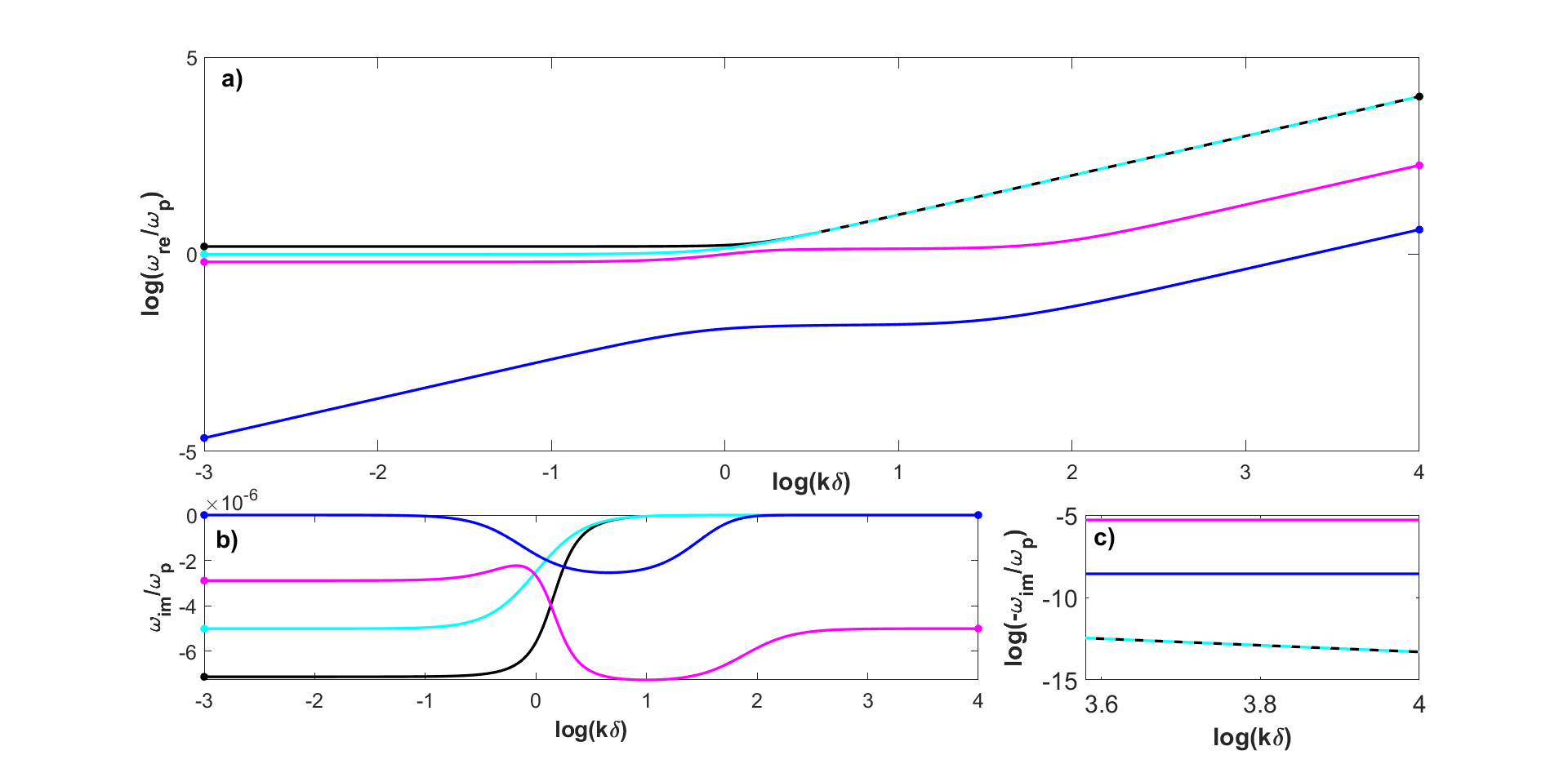}
    \caption{Perpendicular dispersion diagram for $\mu = 1/1836$, $E = 0.935$, $v = 0.018$, $w = 0.0004$ and $\nu = 10^{-5}$. a) Wave frequencies. b) Damping rates. c) Logarithmic scale of damping rates in small wavelength regime. Colouring scheme is the same as in Fig. \ref{fig:coronal-n=1e-8-0g}.
    }
    \label{fig:1-n=1e-5-perp}
\end{figure}

The perpendicular case obtained for $\lambda = 0$ is again similar to the collisionless case where only FMOX branches remain. As expected only these four wave modes are present as neither the slow nor the Alfvén wave propagates in this orientation and the corresponding resonances also vanish for this orientation. The four FMOX remaining modes have a nonzero imaginary part, and their real parts exhibit no new crossings or avoided crossings that were not present in the collisionless case. An illustration of the general behaviour for perpendicular propagation is provided in Fig. \ref{fig:1-n=1e-5-perp}, where we again chose the parameters for a warm plasma in a solar coronal loop, but this time adopted a higher collision frequency, $\nu=10^{-5}$. Note how the damping rates are again in the same order of magnitude for the three MOX cut-off limits, and in the short wavelength limit for the M mode.

\begin{figure}
    \centering
    \includegraphics[width=1.1\textwidth]{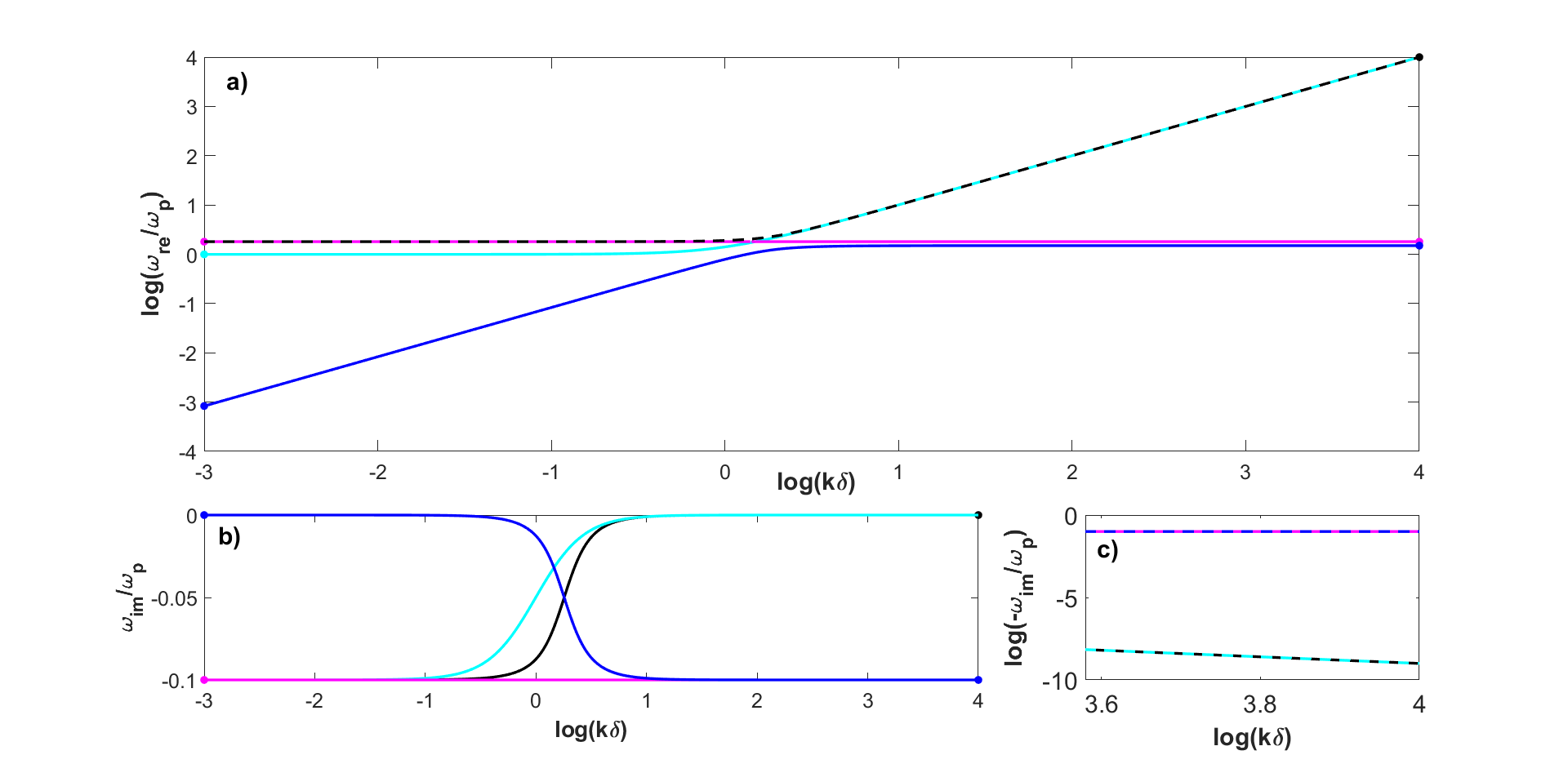}
    \caption{Dispersion diagram for a cold pair plasma, taking $\mu = 1$, $E = 1.5$, $v, w = 0$, $\theta = \pi/2$ and $\nu = 10^{-1}$. a) Wave frequencies. b) Damping rates. c) Logarithmic scale of damping rates in small wavelength regime. Colouring scheme is the same as in Fig. \ref{fig:coronal-n=1e-8-0g}.
    }
    \label{fig:cold-pair-theta=pi_2-nu=1e-1-with-limits}
\end{figure}

For perpendicularly propagating waves in the cold case, a crossing is illustrated between the M and O modes in Fig. \ref{fig:cold-pair-theta=pi_2-nu=1e-1-with-limits}. Note here how the F and M branches are similarly affected by collisions (here $\nu=0.1$), whilst damping also occurs for the three cut-off limits for M, O and X. This crossing was already present in the collisionless case, illustrating that the behaviour of crossings at exactly perpendicular propagation remains unaltered by the collision frequency.

\subsection{Oblique Propagation}

\begin{figure}
    \centering
    \includegraphics[width=1.1\textwidth]{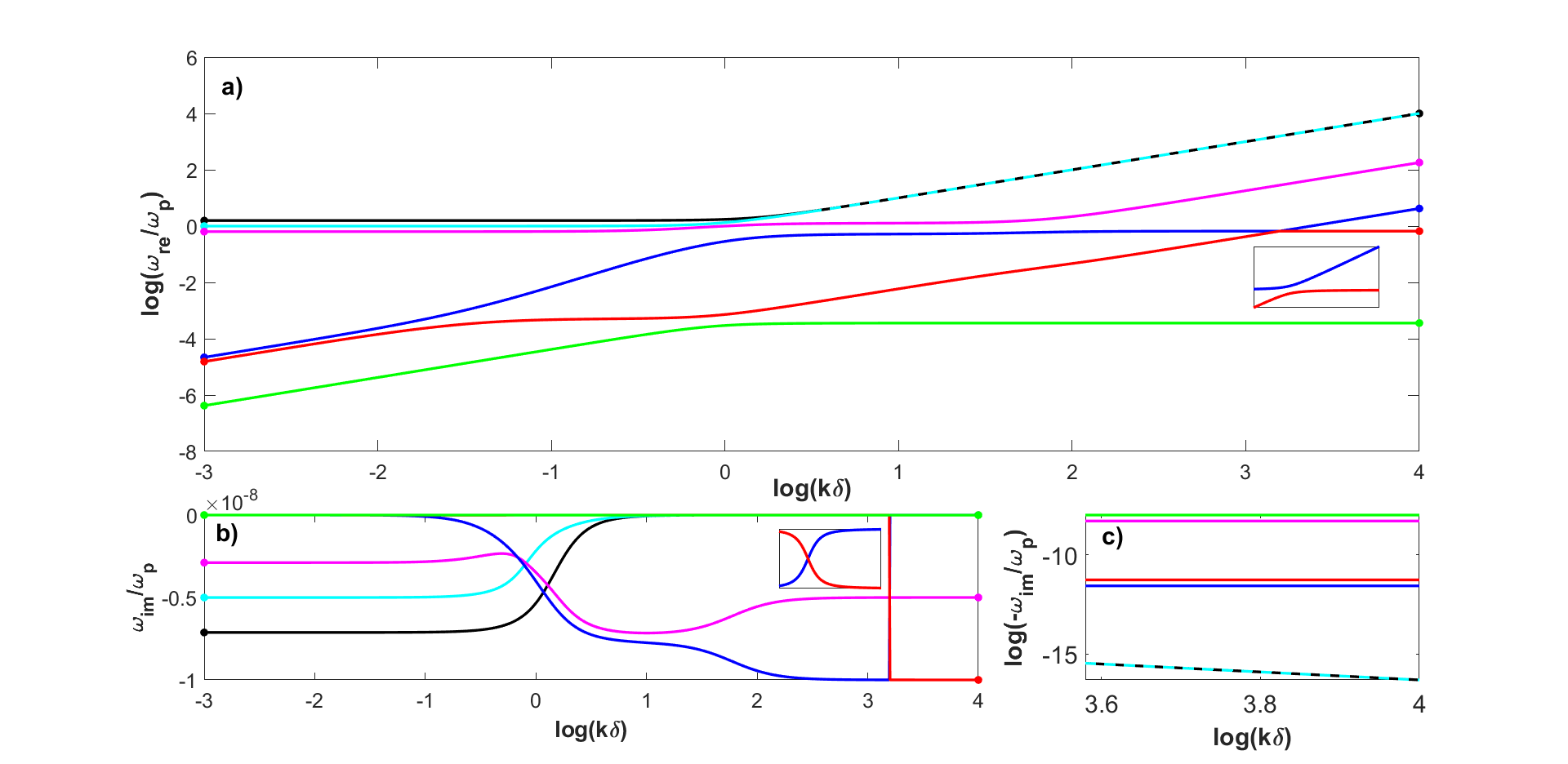}
    \caption{Dispersion diagram for $\mu = 1/1836$, $E = 0.935$, $v = 0.018$, $w = 0.0004$, $\theta = \pi/4$ and $\nu = 10^{-8}$. a) Wave frequencies. b) Damping rates. c) Logarithmic scale of damping rates in small wavelength regime. The insets in panels (a) and (b) highlight the (avoided) crossing behaviour. Colouring scheme is the same as in Fig. \ref{fig:coronal-n=1e-8-0g}.
    }
    \label{fig:coronal-45g-nu=1e-8}
\end{figure}

\begin{figure}
    \centering
    \includegraphics[width=1.1\textwidth]{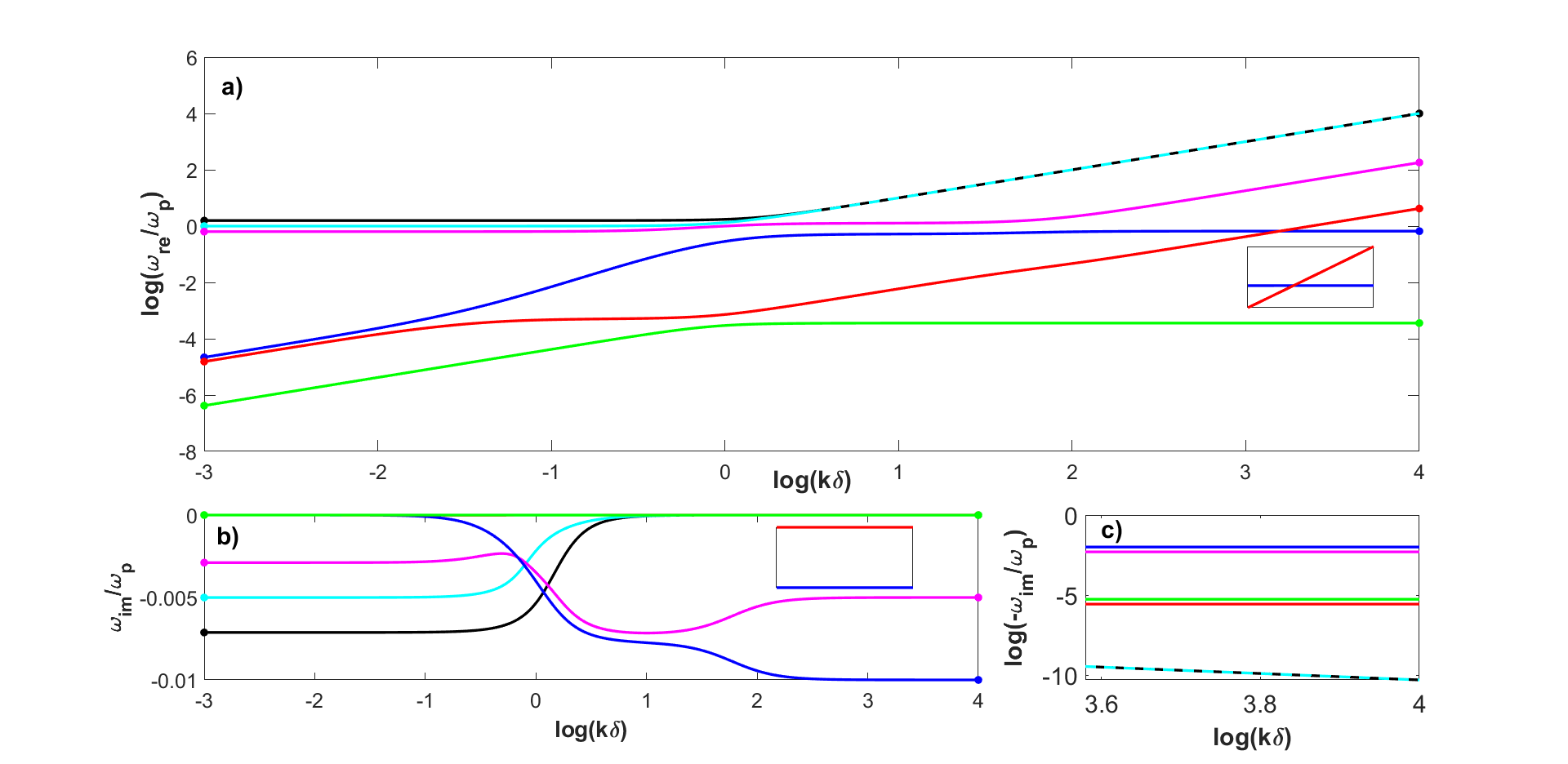}
    \caption{Dispersion diagram for $\mu = 1/1836$, $E = 0.935$, $v = 0.018$, $w = 0.0004$, $\theta = \pi/4$ and $\nu = 10^{-2}$. a) Wave frequencies. b) Damping rates. c) Logarithmic scale of damping rates in small wavelength regime. The insets in panels (a) and (b) highlight the (avoided) crossing behaviour. Colouring scheme is the same as in Fig. \ref{fig:coronal-n=1e-8-0g}.
    }
    \label{fig:coronal-45g-nu=1e-2}
\end{figure}

Finally we turn towards the case of oblique angles between the direction of propagation and the magnetic field lines. Our previous papers \citep{GoedbloedKeppensPoedts2019,Keppens2019a,DeJonghe2020} have demonstrated that in this case an ordering of wave modes exists across all wave numbers, so that 
\begin{center}
    $\omega_\text{S} \leq \omega_\text{A} \leq \omega_\text{F} \leq \omega_\text{M} \leq \omega_\text{O} \leq \omega_\text{X}$.
\end{center}
An initial presumption would be that this is still the case in the collisional regime at least at small values for the collision frequency and could possibly deviate from such behaviour at larger collision frequencies. As a result, both large and small values of $\nu$ were investigated. Two such results are given in Figs. \ref{fig:coronal-45g-nu=1e-8} and \ref{fig:coronal-45g-nu=1e-2}, illustrating that indeed the low collisional case mirrors the collisionless case whereas the more collisional plasma exhibits a true crossing of the wave modes between the A and F modes. (Note that these are the same two modes for which their (avoided) crossing was observed to be angle-dependent in the ideal plasma in \citet{DeJonghe2020}.) These two figures are both for $45^\circ$ angle propagation for the warm solar coronal loop case, but adopt the actual $\nu=10^{-8}$ value in Fig.~\ref{fig:coronal-45g-nu=1e-8}, raised to $\nu=0.01$ in Fig.~\ref{fig:coronal-45g-nu=1e-2}.

\begin{figure}
    \centering
    \includegraphics[width=1.1\textwidth]{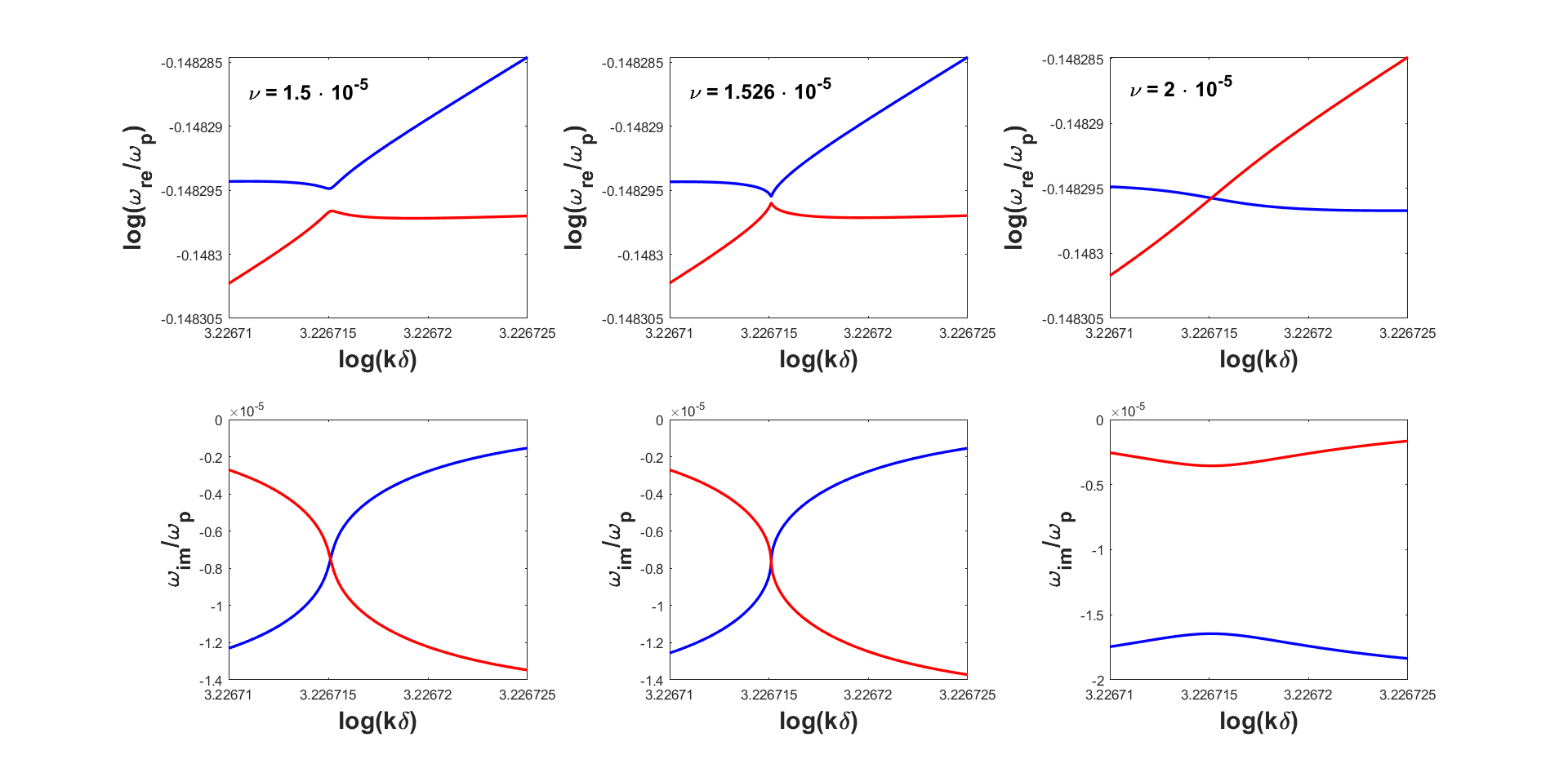}
    \caption{Zoomed in view of the AF crossings and avoided crossings in the dispersion diagram for $\mu = 1/1836$, $E = 0.935$, $v = 0.018$, $w = 0.0004$, $\theta = \sqrt{2}/2$ and varying $\nu$.} Colouring scheme is the same as in Fig. \ref{fig:coronal-n=1e-8-0g}.
    \label{fig:crossings}
\end{figure}

This furthermore suggests that for a given set of plasma parameters and a constant angle, there can be a critical collision frequency determining the transition from avoided to true crossing for a given wave mode pair. The behaviour of the wave modes across such a critical collision frequency is illustrated in Fig. \ref{fig:crossings} with an angle of $\theta \approx 0.7$. This shows that as the collision frequency increases, the real parts approach each other whilst the imaginary parts seem to make a near vertical (but still smooth) jump across the avoided crossing. At the crossing the imaginary parts are very close to each other, but move away as the collision frequency increases until the imaginary parts smooth out. 
It would be of interest to study how the critical collision parameter which turns an AF crossing into an avoided one varies as the plasma and geometric parameters vary, but this is left for future work.

\begin{figure}
    \centering
    \includegraphics[width=1.1\textwidth]{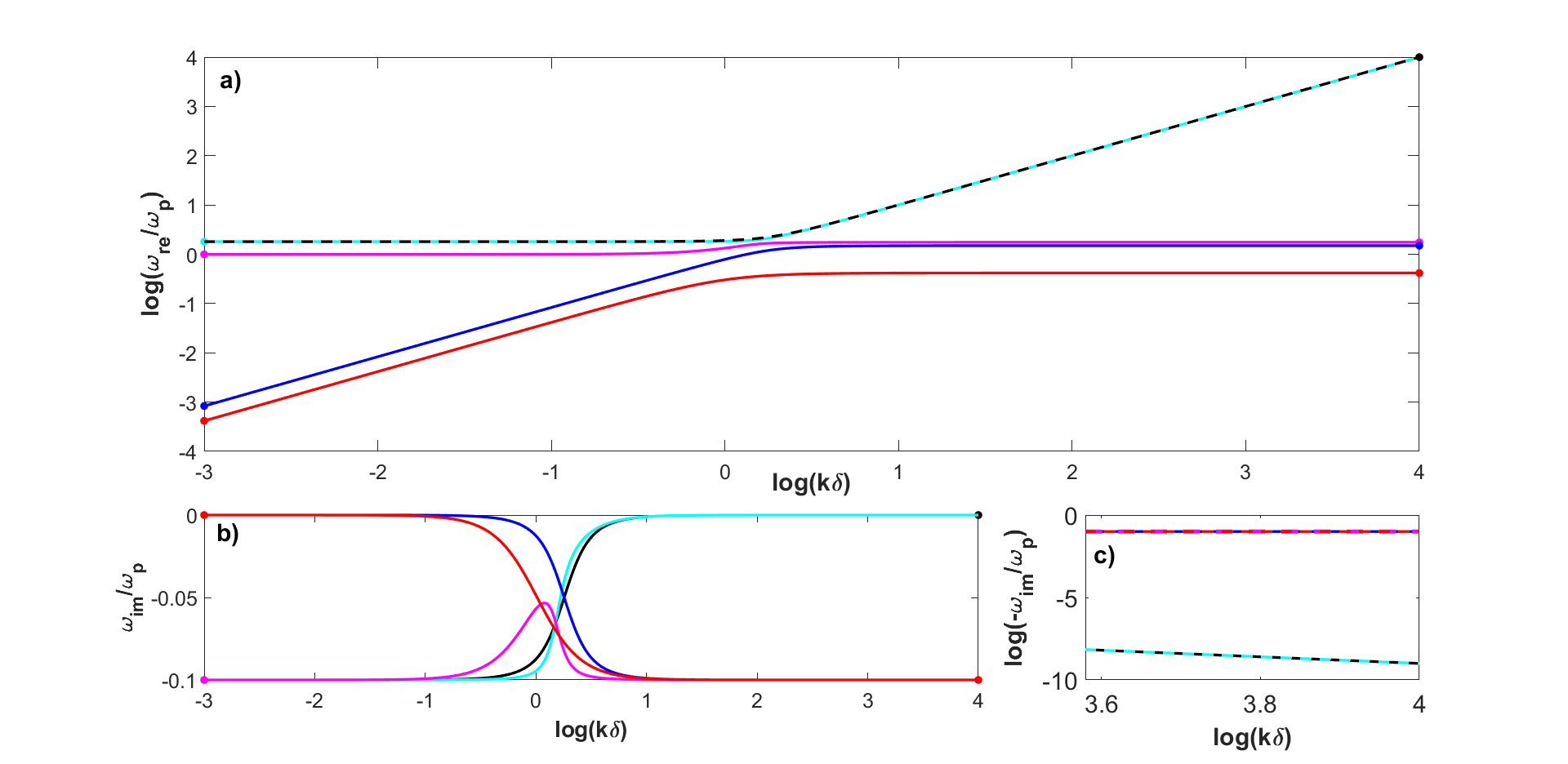}
    \caption{Dispersion diagram for a cold pair plasma, taking $\mu = 1$, $E = 1.5$, $v, w = 0$, $\theta = \pi/3$ and $\nu = 10^{-1}$. a) Wave frequencies. b) Damping rates. c) Logarithmic scale of damping rates in small wavelength regime. Colouring scheme is the same as in Fig. \ref{fig:coronal-n=1e-8-0g}.
    }
    \label{fig:cold-pair-theta=pi_3-nu=1e-1-with-limits}
\end{figure}

\begin{figure}
    \centering
    \includegraphics[width=1.1\textwidth]{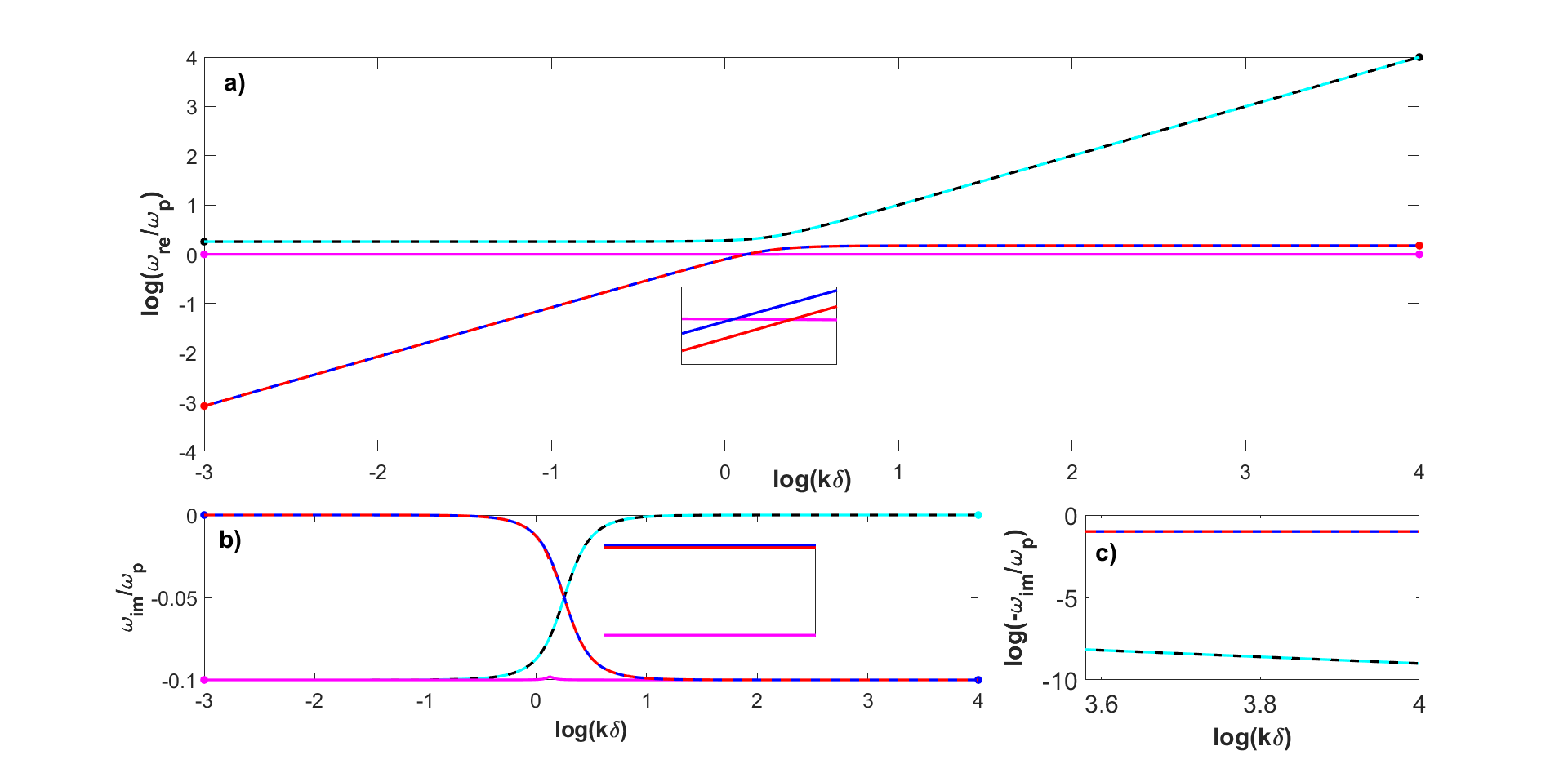}
    \caption{Dispersion diagram for a cold pair plasma, taking $\mu = 1$, $E = 1.5$, $v, w = 0$, $\theta = 0.02$ and $\nu = 10^{-1}$. a) Wave frequencies. b) Damping rates. c) Logarithmic scale of damping rates in small wavelength regime. Colouring scheme is the same as in Fig. \ref{fig:coronal-n=1e-8-0g}.
    }
    \label{fig:cold-pair-theta=0.02-nu=1e-1-E=1.5-with-limits}
\end{figure}

Since the primary result of previous papers was the fact that the behaviour of wave modes depends heavily on the angle, this now raises the question whether varying the angle $\theta \in \left(0,\pi/2\right)$ can turn a barely avoided crossing into a true crossing, or vice versa a true crossing (beyond the critical collision frequency) into an avoided crossing. Hence for constant values of the collision frequency the angle was varied. Fig. \ref{fig:cold-pair-theta=pi_3-nu=1e-1-with-limits} illustrates a cold pair plasma with $\theta = \pi/3$, where no crossings are present. Moving towards the near-parallel case in Fig. \ref{fig:cold-pair-theta=0.02-nu=1e-1-E=1.5-with-limits} with $\theta = 0.02$ and all other parameters fixed, the 5 wave modes become nearly degenerate and two new FM and AM crossings appear. Therefore, for a fixed collision frequency, the dispersion diagram features transitions between crossings and no crossings at oblique angles for variations in the angle.

Since a nonzero collision frequency introduces the possibility of crossings at oblique angles of propagation, the ideal ordering of 
\begin{center}
    $\omega_\text{S} \leq \omega_\text{A} \leq \omega_\text{F} \leq \omega_\text{M} \leq \omega_\text{O} \leq \omega_\text{X}$
\end{center}
is now also violated outside of perfectly parallel or perpendicular propagation. As illustrated by Figs. \ref{fig:cold-pair-theta=pi_3-nu=1e-1-with-limits} and \ref{fig:cold-pair-theta=0.02-nu=1e-1-E=1.5-with-limits}, a crossing at parallel propagation can be maintained at deviations from parallel propagation up to a critical angle, which is determined by the collision frequency and increases as the collision frequency increases. In practice though, realistic collision frequencies will retain this ordering across the bulk of oblique angle values for many plasma environments.

One interesting feature is the fact that the imaginary parts of the cut-offs are clearly organised as the real parts. The imaginary part of the X mode has the highest absolute value and similarly its real part is always above the other wave modes. The ordering of the imaginary parts of the M and O mode however depends then on the ordering of their real parts, or in other words whether or not there are crossings between the O and M mode. As a result we can see that in Fig. \ref{fig:coronal-45g-nu=1e-8} there are no crossings and therefore the ordering in their imaginary parts is $\mathrm{Im}(\omega_\text{X}) < \mathrm{Im}(\omega_\text{O}) < \mathrm{Im}(\omega_\text{M})$ whereas in the presence of crossings such as in Fig. \ref{fig:coronal-n=1e-8-0g} the order becomes $\mathrm{Im}(\omega_\text{X}) < \mathrm{Im}(\omega_\text{M}) < \mathrm{Im}(\omega_\text{O})$. Furthermore, since the O and X modes are light waves in the local high-frequency limit, we expect their imaginary parts to tend to zero, whereas the M mode does have a nonzero damping rate in this limit. This leads to either a crossing of the real parts of the O and M mode or of their imaginary parts. The same is true for the X mode, however since its ordering in the real domain is always above the other wave modes, its real part will never cross the M mode and therefore its imaginary part will always cross the M mode.

\section{Conclusion}\label{sec:conclusion}

Starting from the set of equations describing a two-fluid, ion-electron plasma, the behaviour of the wave modes was analysed in the presence of electron-ion collisions. Due to the introduction of this collision frequency, the wave modes exhibit different damping behaviour based on the value of this collision frequency. For parallel and perpendicularly oriented waves with respect to the background magnetic field, the crossings that were already present in the collisionless case remain unaltered, with only a damping appearing as a result of the collisions. For oblique angles however, the previously avoided crossings remain avoided for collision frequencies sufficiently small and can become true crossings once a critical collision frequency is exceeded. As a result the ordering of the wave modes that was previously found is now no longer valid in general. The behaviour of the damping rates as compared to those obtained from kinetic theory has not been explored in this paper, primarily because $\nu$ was taken as a free parameter. It is left for future work to address proper parametrisations of the collision frequency, to mimic specifically known (kinetic) damping effects on known wave modes.

In future work, the ion-electron model could also be extended with ion-ion and electron-electron collision frequencies. Additionally, damping rates could be determined for the wide variety of whistler waves identified in \cite{DeJonghe2021a} using the ion-electron model to investigate how oblique whistlers are affected by damping. Another aspect not studied here is the actual wave polarisation behaviour along the six branches, which is encoded in the eigenvectors that belong to the specific solutions on each branch. That has recently been done for warm two-fluid settings by \cite{Choi2023}, and this treatise can be followed up with a more comprehensive study of wave damping, polarisation and anisotropic behaviour, as plasma parameters vary.

\begin{acknowledgements}
JDJ and RK acknowledge funding from the European Research Council (ERC) under the European Union Horizon 2020 research and innovation program (grant agreement No. 833251 PROMINENT ERC-ADG 2018). JDJ acknowledges further funding by the UK's Science and Technology Facilities Council (STFC) Consolidated Grant ST/W001195/1. RK is supported by Internal Funds KU Leuven through the project C14/19/089 TRACESpace and an FWO project G0B4521N.\newline

Declaration of Interests. The authors report no conflict of interest.
\end{acknowledgements}

\bibliography{main.bib}

\appendix
\section{Collisional ion-electron dispersion relation}\label{app:disprel}
The dimensionless ion-electron dispersion relation including ion-electron collision frequency $\nu$ can be written as a polynomial of the form
\begin{equation}
    \sum\limits_{\substack{0 \leq m \leq 6\\0 \leq n \leq 4}} \alpha_{mn}\, \omega^{2m} k^{2n} + \imag\nu (1+\mu) \sum\limits_{\substack{0 \leq p \leq 5\\0 \leq q \leq 4}} \beta_{pq}\, \omega^{2p+1} k^{2q} = 0.
\end{equation}
The nonzero coefficients are

\begin{align}
    \alpha_{60} &= 1, \\
    \alpha_{50} &= -\left[3+E^2(1+\mu^2) + 3\nu^2 (1+\mu)^2 \right], \\
    \alpha_{51} &= -(2+v^2+w^2), \\
    \alpha_{40} &= 3+E^2(1+\mu)^2+\mu^2E^4 + \nu^2 (1+\mu)^2 (3+2\mu E^2), \\
    \alpha_{41} &= 4+2E^2(1+\mu^2)+(2+\mu^2E^2+\lambda^2 E^2)v^2+(2+E^2+\lambda^2 \mu^2 E^2)w^2+\cs^2 \nonumber \\ &\quad + \nu^2 (1+\mu)^2 \left[ 2(3+v^2+w^2) - \frac{1-\mu}{1+\mu}(v^2-w^2) \right], \\
    \alpha_{42} &= 1+2v^2+2w^2+v^2w^2, \\
    \alpha_{30} &= -(1+\mu E^2)^2, \\
    \alpha_{31} &= -\bigg\{ 2(1+\mu E^2)^2 + E^2 (1+\lambda^2)(1-\mu+\mu^2) + \left[ 1+\mu^2E^2+\lambda^2 \mu E^2(3+\mu E^2) \right] v^2 \nonumber \\ &\quad + \left[ 1+E^2+\lambda^2 \mu E^2(3+\mu E^2) \right] w^2 + \left[ 2+(1-3\lambda^2)\mu E^2 \right] c_\mathrm{s}^2 \nonumber \\ &\quad + \nu^2 (1+\mu)^2 \left[ 4 +3\cs^2 + \mu E^2 \left( 4 + (1+\lambda^2) \cs^2 \right)\right] \bigg\}, \\
    \alpha_{32} &= -\bigg\{ 1+(1+\mu^2) E^2+2(1+\lambda^2 E^2+\mu^2 E^2)v^2 + 2(1+E^2+\lambda^2 \mu^2 E^2)w^2 + 2c_\mathrm{s}^2 \nonumber \\ &\quad + \left[ 2+\lambda^2E^2(1+\mu^2) \right] v^2w^2 + \nu^2 (1+\mu)^2 \left( 3 +2v^2 +2w^2 +v^2w^2 +4\cs^2 \right) \bigg\}, \\
    \alpha_{33} &= -(v^2+w^2+2v^2w^2), \\
    \alpha_{21} &= \mu E^2(1+\mu E^2)(1+\lambda^2) + (1+\mu E^2)(1+\lambda^2 \mu E^2)c_\mathrm{s}^2, \\
    \alpha_{22} &= \mu E^2 (1+\mu E^2) + \lambda^2 E^2(1-\mu +\mu^2) + \left[ (1+\lambda^2)\mu^2 E^2 + 2\lambda^2 \mu E^2(2+\mu E^2) \right]v^2 \nonumber \\ &\quad + \left[ (1+\lambda^2)E^2 + 2\lambda^2 \mu E^2(2+\mu E^2) \right]w^2 + \left[ 2+(1-5\lambda^2)\mu E^2 \right]c_\mathrm{s}^2 \nonumber \\ &\quad + (1+\lambda^2 \mu E^2)^2 v^2w^2 + \nu^2 (1+\mu)^2 \left\{ 1 +4\cs^2 +2\mu E^2 \left[ 1 +(1+\lambda^2) \cs^2 \right] \right\}, \\
    \alpha_{23} &= (\mu^2E^2+\lambda^2 E^2)v^2 + (E^2+\lambda^2 \mu^2E^2)w^2 + c_\mathrm{s}^2 + 2\left[ 1+\lambda^2 E^2(1+\mu^2) \right] v^2w^2 \nonumber \\ &\quad + \nu^2 (1+\mu)^2 (v^2 +w^2 +2v^2w^2 +2\cs^2), \\
    \alpha_{24} &= v^2w^2, \\
    \alpha_{12} &= -\lambda^2 \mu E^2 \left\{ \mu E^2 + \left[ 2+\mu E^2(1+\lambda^2) \right] c_\mathrm{s}^2 \right\}, \\
    \alpha_{13} &= -\lambda^2 \left\{ \mu^2E^4(v^2+w^2) + E^2(1+\mu^2)c_\mathrm{s}^2 + 2\mu E^2 (1+\lambda^2 \mu E^2) v^2w^2 \right\} \nonumber \\ &\quad - \nu^2 (1+\mu)^2 \cs^2 \left[ 1 + \mu E^2 (1+\lambda^2) \right], \\
    \alpha_{14} &= -\left[ \lambda^2 E^2(1+\mu^2) + \nu^2 (1+\mu)^2 \right] v^2w^2, \\
    \alpha_{03} &= \lambda^4 \mu^2E^4 \cs^2, \\
    \alpha_{04} &= \lambda^4 \mu^2 E^4 v^2w^2, \\
    &\text{and} \nonumber \\
    \beta_{50} &= 3, \\
    \beta_{40} &= - \left[ 6+(1+\mu)^2 (E^2+\nu^2) \right], \\
    \beta_{41} &= -\left( 6 +2v^2 +2w^2 +\cs^2 \right), \\
    \beta_{30} &= (1+\mu E^2) (3+\mu E^2), \\
    \beta_{31} &= (1+\mu)^2 (E^2+\nu^2) (2+\cs^2) + \mu E^2 (3\lambda^2-1) \left[ \cs^2 + \frac{1-\mu}{1+\mu} (v^2-w^2) \right] \nonumber \\ &\hspace{2cm} + 2\left[ 2(2+v^2+w^2) - \frac{1-\mu}{1+\mu} (v^2-w^2) \right], \\
    \beta_{32} &= 3 +4v^2 + 4w^2 +2v^2w^2 +2\cs^2, \\
    \beta_{21} &= -\left\{ 2 +3\cs^2 +\mu E^2 \left[ 5 +2\cs^2 +\lambda^2 (1+2\cs^2) \right] +\mu^2E^4 (2 +\lambda^2\cs^2) \right\}, \\
    \beta_{22} &= -\bigg\{ 2(1 +v^2 +w^2 +v^2w^2 +3\cs^2) + (1+\mu)^2 (E^2+\nu^2) (1+2\cs^2) \nonumber \\ &\quad  +2\mu E^2 \left[ \lambda^2v^2w^2 + (3\lambda^2-1)\left( \cs^2 + \frac{1-\mu}{1+\mu} (v^2-w^2) \right) \right] \bigg\}, \\
    \beta_{23} &= -(2v^2 +2w^2 +4v^2w^2 +\cs^2), \\
    \beta_{12} &= 2\cs^2 +\mu E^2 \left[ 1+2\cs^2 + \lambda^2 (1+4\cs^2) \right] +\mu^2 E^4 (1+2\lambda^2 \cs^2), \\
    \beta_{13} &= 2v^2w^2 +2\cs^2 + (1+\mu)^2 (E^2+\nu^2) \cs^2 \nonumber \\ &\quad + \mu E^2 \left[ 4\lambda^2 v^2w^2 + (3\lambda^2-1) \left( \cs^2 + \frac{1-\mu}{1+\mu} (v^2-w^2) \right) \right], \\
    \beta_{14} &= 2v^2w^2, \\
    \beta_{03} &= -\lambda^2 \mu E^2 \cs^2 (2+\mu E^2), \\
    \beta_{04} &= -2\lambda^2 \mu E^2 v^2w^2.
\end{align}

\bibliographystyle{jpp}

\end{document}